\documentclass[aps, prd, amsmath, floats, floatfix, twocolumn,superscriptaddress, nofootinbib, showpacs,reprint]{revtex4}

\usepackage{hyperref}
\usepackage{graphicx}
\usepackage{amsmath,amssymb}
\usepackage{amsfonts}
\usepackage{xspace} 
\usepackage[usenames]{color}
\usepackage{dcolumn}
\usepackage{bm}
\usepackage{mathrsfs}

\usepackage{ulem}
\definecolor {darkgreen}{rgb}{0.2,0.7,0.2}

\newcommand{\eq}{\begin{equation}}
\newcommand{\be}{\begin{equation}}
\newcommand{\eeq}{\end{equation}}
\newcommand{\ee}{\end{equation}}

\begin{document}

\title{Constraining scalar-tensor theories of gravity from the most massive neutron stars}

\author{Carlos Palenzuela}
\affiliation{Departament de F\'isica, Universitat de les Illes Balears and Institut d'Estudis Espacials de Catalunya, Palma de Mallorca, Baleares
E-07122, Spain}
\author{Steven L. Liebling}
\affiliation{Department of Physics, Long Island University, Brookville, New York 11548, USA}

\begin{abstract}

Scalar-tensor~(ST) theories of gravity are natural phenomenological extensions to general relativity.
Although these theories are severely constrained both by solar system experiments and by binary pulsar observations,
a large set of ST families remain consistent with these observations. Recent work has suggested
probing the unconstrained region of the parameter space of ST theories based on the stability properties
of highly compact neutron stars.
Here, the dynamical evolution of very compact stars in a fully nonlinear code
demonstrates that the stars do become unstable 
and that the instability, in some cases, drives the stars to collapse.
We discuss the implications of these results in light of recent observations of the most
massive neutron star yet observed.
In particular, such observations suggest that such a star would be subject to 
the instability for a certain regime; its existence therefore supports a bound on the ST parameter space.
\end{abstract}

\date{\today \hspace{0.2truecm}}


\maketitle

\noindent\textbf{\textit{Introduction:}}
%
Despite the tremendous success of general relativity throughout its first hundred years
as a description of gravity, alternative theories remain attractive. Motivated either by
purely theoretical ideas such as string theory or more phenomenological issues such as dark
matter~\cite{Brito:2015yga}, alternative gravity is severely constrained
by observations~\cite{2014LRR....17....4W}.
One particularly interesting class of theories are the 
scalar-tensor~(ST)
theories of gravity in which gravity is mediated by a metric and a scalar
field. 
A primary advantage of ST theories is that they 
retain a characteristic structure similar to that of GR and are similarly well-posed. Another advantage is that, in an appropriate limit, GR is recovered. 
Hence, ST theories will always have some parameter region that satisfies the same observational constraints obeyed by GR.
Nevertheless, ST theories offer additional degrees of freedom with which to explore phenomena such as dark energy.

If the universe allows for such degrees of freedom, then perhaps we can find observational
consequences using the population of observed neutron stars~(NS).
Here we restrict ourselves to a particular set of ST theories, namely the Damour-Esposito-Farese~\cite{1993PhRvL..70.2220D,1996PhRvD..54.1474D} model which is characterized by two constants, 
 $\{\varphi_0, {       \beta} \}$.
The first constant, $\varphi_0$, is the asymptotic value of the scalar field which is constrained
by solar system observations to be quite small. The second constant, $\beta$, measures the linear,
 effective 
coupling between the scalar field and the regular matter content (more detail and the precise form of the action follows below). When both these constants vanish, one recovers GR.

Previous work with neutron stars found for $\beta < -4.5$ that stars underwent spontaneous scalarization in which the scalar field acquired a value in the neighborhood of the star much larger than its
asymptotic value $\varphi_0$.  As discussed below, scalarization effects lead to
constraints on $\beta$ from below.
However, the positive $\beta$ region remains largely unconstrained, and it is therefore important
to find dynamics in this region with which to compare to observations. 

For $\beta$ above some critical value, $\beta_{\rm crit}$,
an instability for very compact neutron stars has been found~\cite{2015PhRvD..91f4024M}.
Although 
noticed quite some time ago (i.e., see for instance Fig.~5 in Ref.~\cite{1997PThPh..98..359H} and also in Ref.~\cite{2011PhRvD..83h1501P}), it was only recently realized
that this instability could appear in the stable branch of equilibrium
configurations of realistic microphysical EoS. 
Although linear perturbation analysis reveals the instability, such an analysis does not indicate the
end-state or whether the instability could simply drive the system to a stable configuration.
This instability may occur in a wider class of  alternative gravity theories, such as $f(R)$ gravity~\cite{Capozziello:2015yza}
(also see Section~IIIA of Ref.~\cite{Sotiriou:2008rp}),
and so similar similar bounds may be found for other theories.

In order to study this instability and its end state, 
we perform fully non-linear dynamical evolutions of neutron star solutions with different compactness $C \equiv G M/R c^2$
until they relax to their final stationary state. 
For stars with low compactness ($C< 0.27$), the initial data simply relaxes to a ST
equilibrium configuration for any value of $\beta$. For very compact stars ($C\ge 0.29$)
the instability drives the star to collapse to a black hole if $\beta$ exceeds some critical value, $\beta_{\rm crit}\approx 90$. 
For intermediate compactness ($C\approx 0.28$), these two disparate behaviors are also observed, but
another behavior, intermediate between collapse and stability, appears. For $\beta>\beta_{\rm crit}$ but $\beta$ not too large,
the central scalar field grows promptly until the instability saturates and then the solution undergoes damped oscillations about
an ST equilibrium solution.

In addition to studying this instability in the fully nonlinear regime, we also
are able to study non-spherically symmetric stars which allow for rotation. For rotating
stars, we find similar behavior of the instability.
We discuss the implications of these results
in the astrophysical context of very massive neutron stars.
The most massive neutron star ever observed has mass $M_{\rm NS}= 2.01 \pm 0.04 M_{\odot}$~\cite{2013Sci...340..448A} and an estimated radius of roughly $R_{\rm NS} \approx 10.5^{+ 1.2}_{- 1.0}$km~\cite{2015arXiv150505155O}. 
These values lead to an estimated  compactness high enough ($C\approx 0.28$) under most EoS to
develop this instability, and therefore the existence of the star suggests an upper bound of
$\beta \approx \mathcal{O}(1000)$.

~\\
\noindent\textbf{\textit{Phenomenological constraints:}}
We consider a general ST theory with action
\begin{equation} \label{Jframe_action}
S=\!\!\int d^4 x\frac{\sqrt{-g}}{2\kappa}\left[\phi R-\frac{\omega(\phi)}{\phi} \partial_\mu \phi \partial^\mu \phi
\right]+ S_M[g_{\mu\nu},\psi]\,,
\end{equation}
where  $\kappa=8 \pi G$ (adopting $c=1$ throughout this paper),
$R$ is the Ricci scalar,  $g$ is the determinant of the metric,
$\phi$ is the gravitational scalar, 
and $\psi$ collectively describes the matter degrees of freedom.
Although Eq.~\eqref{Jframe_action} is not the most general action yielding second-order field equations, it includes a large set of theories by allowing the coupling $\omega(\phi)$ to depend on the scalar fields. Notice that, for simplicity, no  potential $V(\phi)$ is included here.

One can re-express the (``Jordan-frame'') action \eqref{Jframe_action} in
the so-called ``Einstein-frame'' through a conformal transformation $g^E_{\mu\nu}=\phi\, g_{\mu\nu}$, which yields
\begin{equation}
\label{einframe}\!
\!S=\!\!\int\!d^4 x \sqrt{-g^E} \left( \frac{R^E}{2\kappa}\!-\!\frac{1}{2}g_E^{\mu\nu} \partial_\mu\varphi \partial_\nu\varphi
\right) \! +S_M\!\!\left[\frac{g^E_{\mu\nu}}{\phi(\varphi)},\psi\right]\!\!\!
\end{equation}
where this scalar field, $\varphi$, is defined in terms of $\phi$ by
$({{\rm d}\log \phi}/{{\rm d}\varphi})^2\equiv {2\kappa}/[{3+2 \omega(\phi)}]$.
A common feature of these families of ST theories is the absence of a direct coupling between the matter degrees of freedom and the scalar field, since otherwise there would be a scalar ``fifth force'' that has not been
observed. However, there is an indirect coupling between matter and scalar field mediated by gravity, which can be expanded in Taylor series around the
asymptotic ``vacuum'' value of the scalar field, $\varphi_0$. For instance, 
the Jordan-Fierz-Brans-Dicke (JFBD) theories are recovered by considering only the first constant term in this expansion, although solar system tests
constrain this constant to very small values.

In this work we focus on the Damour and Esposito-Farese (DEF) family of ST theories~\cite{1993PhRvL..70.2220D,1996PhRvD..54.1474D} which are the next most simple family. In DEF theories, the coupling includes up to the term linear in $\varphi$ in the Taylor series 
expansion 
\begin{equation}\label{coupling_expansion_dimension}
     \frac{1}{2} \frac{{\rm d}\log \phi}{{\rm d}\varphi}  =  - (4 \pi G) \beta \varphi  
\end{equation}
corresponding to the choice
\begin{equation}
   \omega(\phi)=-3/2- 1/(2 \beta \log\phi) \,\, ,\,\,\, \phi=\exp(- 4 \pi G \beta \varphi^2).
\end{equation}

The DEF theories, characterized by the coupling constant $\beta$ and 
by the asymptotic value of the scalar field $\varphi_0$, have been 
shown to produce effects differing significantly from GR in strong-field regimes such as 
the interior of NS~\cite{1993PhRvL..70.2220D,1996PhRvD..54.1474D}. 
These effects have led to a number of observational constraints.
More specifically, if $\beta$ is sufficiently negative,
the trivial vacuum of the scalar field becomes unstable and it 
becomes energetically favorable for the scalar field to settle down 
into a non-trivial configuration inside the NS (``spontaneous scalarization''). 
This configuration with a
large scalar field in the stellar interior affects not only NS mass and its radius, 
but also its orbital motion if the star is in a binary.  
This modification to orbital motion arises because scalarization enhances the 
gravitational attraction between the components of the binary and triggers 
the emission of dipolar scalar 
radiation~\cite{1975ApJ...196L..59E,1992CQGra...9.2093D,1989ApJ...346..366W}. 
These effects lead to constraints on the nondimensional constant $\beta$ such that 
$\beta \gtrsim -4.5$ using binary pulsar data~\cite{1996PhRvD..54.1474D,1998PhRvD..58d2001D,2012MNRAS.423.3328F,2013Sci...340..448A} while the Cassini experiment constrains $\varphi_0< 1.26 \times10^{-2}G^{1/2}/|\beta|$. 

All these results, however, probe the strong-field, mildly relativistic regime since they involve either static/stationary configurations or velocities much smaller than the speed of light (e.g. binary pulsars have velocities $v\approx 10^{-3} c$). 
Simulations of coalescing binary neutron stars show that large deviations from
 GR develop at separations much smaller than those observed with binary pulsars,
providing signals that are, at least in principle, observable with existing gravitational-wave detectors  such as Advanced LIGO/Virgo~\cite{2013PhRvD..87h1506B,2014PhRvD..89d4024P,2014PhRvD..89h4005S}. 

Much less constrained is the positive regime of $\beta$ (see e.g. Fig.~7 of Ref.~\cite{2013Sci...340..448A}). Indeed, cosmological studies
have shown that in the $\beta > 0$ regime ST theory approaches GR exponentially (i.e. GR is an attractor in the theory
phase space) and that the parametrized post-Newtonian~(PPN) parameters are exponentially close to their GR
values~\cite{PhysRevLett.70.2217,PhysRevD.48.3436}.
Thus the instability studied in Ref.~\cite{2015PhRvD..91f4024M} represents
an important opportunity to constrain the positive $\beta$ regime.
A significant observation of Ref.~\cite{2015PhRvD..91f4024M}
is that the high compactness required by this instability could occur
for otherwise stable NS (ie with masses less than the maximum supported by
the EoS).
Here we therefore evolve high compactness stars within DEF gravity to
study nonlinear effects and the ultimate fate of stars subject to this instability.

~\\
\noindent\textbf{\textit{Equations of Motion:}\label{theory_motion}}
In the Einstein frame the ST field equations are quite similar to the standard Einstein
equations minimally coupled to a scalar field but with 
a coupling between the scalar field and the matter. The full system
of equations takes the form
\begin{gather}\label{einstein}
   G_{\mu\nu}^{E}=\kappa \left(T^\varphi_{\mu\nu} + T^E_{\mu\nu} \right),\\
   \Box^E \varphi = - (4 \pi G) \beta \varphi T_E\label{KG} \,,\\
   \nabla_\mu^E T_E^{\mu\nu}= (4 \pi G) \beta \varphi T_E 
   g_E^{\mu\nu} \partial_\mu \varphi\,,\label{scalarT}   
\end{gather}
where
\begin{eqnarray} \label{Tfluid}
T_E^{\mu\nu}&=& [{\rho}_E (1+ \epsilon_E) + p_E ] u_E^\mu u_E^\nu + p_E g^E_{\mu\nu} \,\,\,\,\, \mbox{and} \\
\label{Tscalarfield}
T^\varphi_{\mu\nu} &=&\partial_\mu \varphi \partial_\nu \varphi- \frac{g^E_{\mu\nu}}{2} g_E^{\alpha\beta} \partial_\alpha \varphi
\partial_\beta \varphi\, 
\end{eqnarray} 
are the matter and scalar-field stress-energy tensors in the Einstein frame
and $T_E\equiv T_E^{\mu\nu}g^E_{\mu\nu}$.
Baryon number conservation in the Jordan frame leads to
\eq\label{rest_mass_cons_Einstein}
\nabla^E_\mu j_E^\mu = 
(4 \pi G) \beta \varphi j_E^\mu \partial_\mu \varphi \,,
\eeq
with $j_E^\mu={\rho}_E u_E^\mu$.
In other words, neither the current associated with baryon number
nor the stress-energy tensor are
generally conserved in the Einstein frame and instead has source
terms.
Solving the system~\eqref{einstein},~\eqref{KG},~\eqref{scalarT} and~\eqref{rest_mass_cons_Einstein}
and transforming back to the Jordan frame provides a solution to the original problem.

~\\
\noindent\textbf{\textit{Initial data:}\label{initial_data}}
%
We construct the initial data for our neutron stars as follows.
We begin by solving for  a stellar configuration that is at equilibrium in GR as described below. 
In addition, we specify the initial scalar field as a constant, $\varphi(x^i,t=0) = \varphi_0$.

\begin{figure}
\centering
\includegraphics[width=8.8cm]{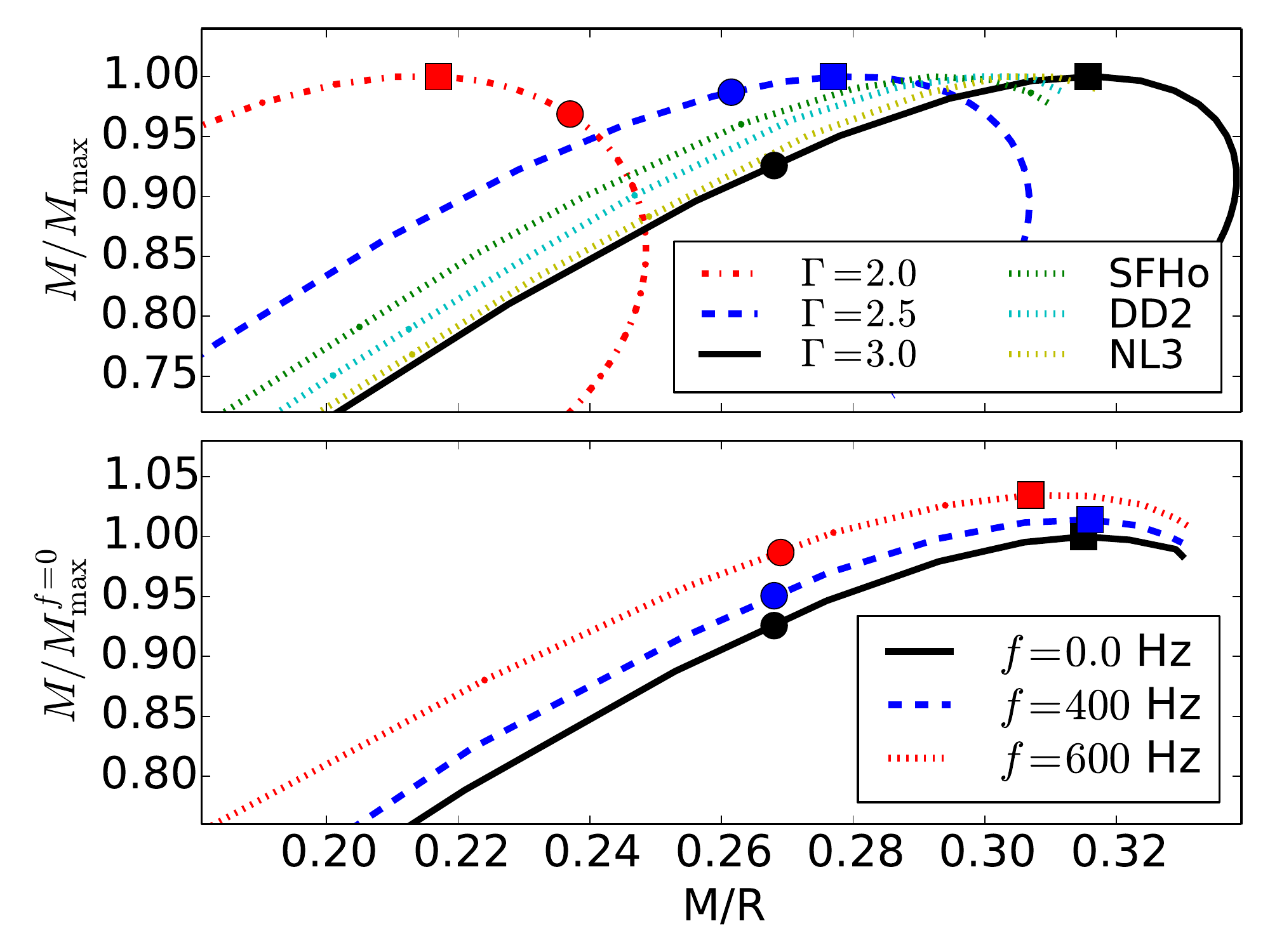} 
\caption{Characterization of initial stellar models.
\textbf{Top:}
Mass versus compactness of spherical, non-rotating neutron star equilibrium configurations for different polytropic exponent (e.g. stiffness) in
the polytropic EoS $p=K \rho^{\Gamma}$. The transition between models with $T$ everywhere negative to models with $T$ positive in a central region is marked 
(solid circles) for each value of $\Gamma$. The stable branch lies on the 
left of the maximum of the curve. Solutions for three microphysical EoS are also 
included, showing that their structure lies between that of the polytropes with
$\Gamma=2.5$ and $\Gamma=3$.
\textbf{Bottom:}
Mass versus compactness for rigidly rotating neutron stars with fixed $\Gamma=3$ but different rotational frequencies. Notice that the transition
 in $T$ lies at roughly the same compactness as for the nonrotating models.
However, as the frequency increases, the configuration at the transition has
a mass closer to the maximum mass than the nonrotating case.
Again, the stable branch lies on the left of the maximum mass. 
\label{fig:stellarsolutions}
}
\end{figure}

It is straightforward to show that these initial data are also solutions to ST theory
in the Einstein frame. In particular, the stress-energy tensor of the scalar field, 
$T^\varphi_{\mu\nu} $,
given by Eq.~(\ref{Tscalarfield}) is initially zero when $\varphi$ is constant, and 
hence Eq.~(\ref{einstein}) is satisfied by the GR solution. The fluid equations in
 Eq.~(\ref{scalarT}) are also satisfied  by a constant scalar field. 
Although the initial data is a solution within ST theory, the scalar field is
sourced via Eq.~\eqref{KG} and hence is no longer in equilibrium. 
Instead, the initial data is at a moment of time symmetry
with an initial ``acceleration.''

These solutions represent perturbed stellar configurations, and this same procedure is adopted
both by Refs.~\cite{2015PhRvD..91f4024M} and ~\cite{1997PThPh..98..359H} in their linear perturbation analysis.
In this case, the magnitude of the perturbation depends, in part, on the value of $\beta$.
Indeed, one could imagine a scenario in which such a perturbed star might result naturally. 
In particular, consider a neutron star with small compactness, so that the star begins with
a small, nearly constant scalar field in its interior. The compactness might increase suddenly
by going supernova, accreting mass, or undergoing a merger, while maintaining a fairly constant and small scalar field.

We use the {\sc Lorene} code~\cite{lorene} to construct the initial stellar models.
The stars are described by a polytropic equation of state, $p/c^2=K \rho_0^\Gamma$, which is a
reasonable approximation for cold stars.  
In our search for solutions subject to the instability, 
we explore a range of values of the adiabatic index $\Gamma=2-3$, but only $\Gamma=3$ is
chosen for our evolutions.
The choice here of $K$ is not important since the solution can be rescaled to match the maximum mass $M \approx 2.0 M_\odot$ of the most massive observed neutron star~\cite{2013Sci...340..448A}.

We focus on configurations fulfilling two conditions: 
(i)~lying on the stable branch of solutions and 
(ii)~having compactness sufficient to achieve a region with $T > 0$ in its central interior (in contrast, 
stars generally have $T<0$ everywhere). This second condition is crucial for the
existence of the instability, and we have verified that no instability is apparent for
configurations which lack such a region. Notice that the positivity
of the trace is independent of the frame because $T = \phi^2 T_E$.
In the case of the perfect fluid described by Eq.~(\ref{Tfluid}), the trace can be written as
\begin{equation}
  T = 3 p - \rho (1 + \epsilon) 
    = \left( \frac{3\Gamma-4}{\Gamma-1} \right) p - \rho
    = \left( \frac{3\Gamma-4}{\Gamma-1} \right) K \rho^{\Gamma} - \rho  
\end{equation}
where the first equality is completely generic and in the second we have used the ideal gas EoS law, $p = (\Gamma -1) \rho \epsilon$.
The third equality uses a polytropic relation for the pressure which only applies to isentropic fluids without shocks.

We characterize a range of these solutions in Fig.~\ref{fig:stellarsolutions}.
The mass of nonrotating solutions is shown in the top panel of Fig.~\ref{fig:stellarsolutions} as
a function of compactness. Where the maximum mass is achieved represents a turning point (marked
with a square)
in stability such that solutions of smaller compactness are stable (in GR) whereas more compact solutions
are unstable. Three families of solutions, each with a different EoS parameter $\Gamma$, are shown,
and it is apparent that larger values correspond to more compact stars.

As these stars get more compact, the solutions eventually reach a point where $T$ becomes
positive in their interior. The point at which solutions achieve $T=0$ is marked with a circle
so that more compact stars have regions with positive trace. It is important to note that
for small $\Gamma$, the circle is to the right of the square whereas for large $\Gamma$ one has
the opposite. This note is important because, for $\Gamma=3$ for example, one has solutions
on the stable branch that nevertheless have regions of positive trace. It is precisely the evolution
of these solutions that is of interest here.
Interestingly, the null point seems mostly independent of the value of $\Gamma$, lying at a compactness
of $C \approx 0.27$. Indeed, the results of Ref.~\cite{2015PhRvD..91f4024M} with
realistic EoS also find similarly high compactness solutions on the stable branch. We include the family of
curves of other realistic EoS in Fig.~\ref{fig:stellarsolutions} showing that these EoS have solutions
similar to that of $\Gamma=3$ despite spanning a wide range of stiffness (e.g. resulting radii). 

A similar plot but instead for rigidly rotating solutions is shown in the bottom panel of
 Fig.~\ref{fig:stellarsolutions}.
Here, instead of different values of $\Gamma$, solutions rotating at different frequencies, $f$, are shown,
all for $\Gamma=3$.
As before, the separation between stability branches is marked with a square and
the null point where $T$ changes sign is marked with a circle. Notice that the null point is roughly at the same compactness $C \approx 0.27$ for any  frequency.

~\\
\noindent\textbf{\textit{Numerical simulations:}\label{simulations}}
%
Our numerical code for evolving these configurations 
has been described and tested previously~\cite{2007PhRvD..75f4005P,2008PhRvD..77b4006A}
finding a dynamical scalarization effect in
the evolution of binary NS systems in ST~\cite{2013PhRvD..87h1506B,2014PhRvD..89d4024P}. 
The initial data are evolved on a cubical computational domain 
$x^i \in [-160,160]$ employing fixed mesh refinement with a finest grid spacing $\Delta x = R_{\rm NS}/50$.
Certain evolutions were conducted at higher resolutions
up to $\Delta x = R_{\rm NS}/70$ to verify convergence.
We vary the
effective coupling by considering different values of $\beta$
(i.e., keeping $\varphi_0\leq 10^{-5} G^{-1/2}$ fixed). We have checked that the results do not change significantly when $\varphi_0$ is varied within those bounds.  

We employ an ideal gas EoS for our simulations choosing $\Gamma=3$
so that solutions containing $T>0$ lie on the stable branch in GR. We evolve initial data
along this family parametrized by the compactness.

We begin by discussing our evolutions of a configuration with compactness $C=0.26$.
For this initial data, the trace is everywhere negative and no instability is expected.
Indeed, in our evolutions we increased $\beta$ up to a value of $8000$ with no indication of instability.
In particular, the central density and lapse remained unchanged with increasing $\beta$.
The scalar field demonstrated fast oscillations with an amplitude damped in time. For very
large values of $\beta$, the overall amplitude was diminished. 

\begin{figure}
\centering
\includegraphics[width=8.8cm]{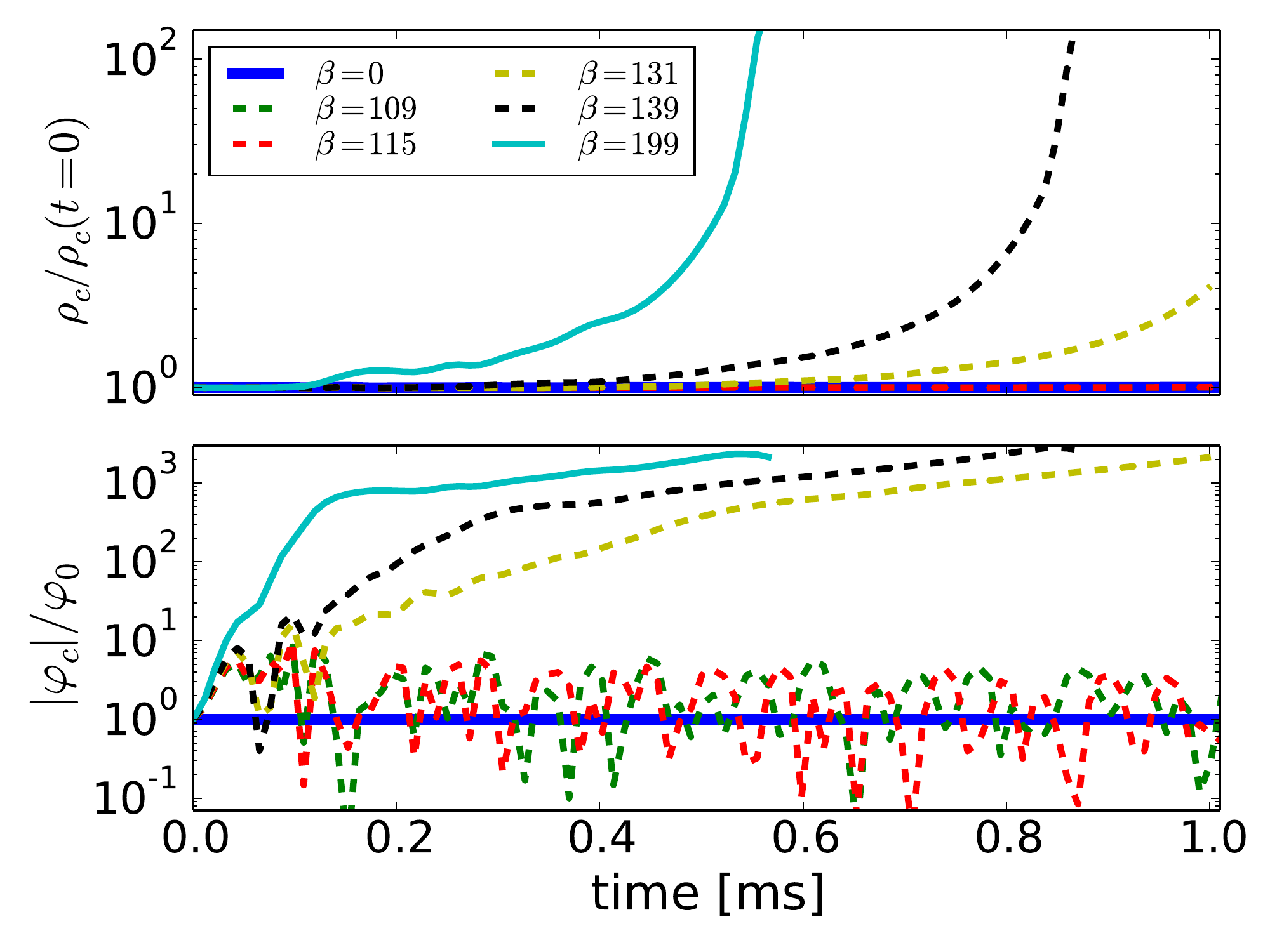}
\caption{{\it Dynamics of a highly compact ($C=0.29$), non-rotating star}.
\textbf{Top:}~Central value of the rest-mass density, $\rho_c$,  as a function of time.
\textbf{Bottom:}~Central value of the scalar field, $\varphi_c$.
Note that the $\beta=0$ solution is stable whereas the scalar field grows initially
for nonvanishing $\beta$. However, only for $\beta>\beta_{\rm crit}$ 
does the central density grow apparently without bound, collapsing to a black hole.
For $\beta < \beta_{\rm crit}$, the scalar field oscillates, sometimes 
transitioning to negative values.
Here $\beta_{\rm crit} \approx 90\pm 10$ with the uncertainty arising due to late
time instability.
\label{fig:C029_central_values}}
\end{figure}

In stark contrast, the evolution of a high compactness configuration ($C=0.29$) is shown
in Fig.~\ref{fig:C029_central_values}. These evolutions display two disparate behaviors
depending on the value of $\beta$. For $\beta>\beta_{\rm crit}$~(here $\beta_{\rm crit} \approx 90$), the central density and
scalar field grow more than an order of magnitude while the star collapses to a black hole.
For $\beta<\beta_{\rm crit}$, these quantities oscillate (but with a slow drift that decreases with increasing resolution). In particular, the
density slowly increases by a few percent while the lapse decreases by an even smaller amount. 
The scalar field oscillates about either zero or its initial value. 

\begin{figure}
\centering
\includegraphics[width=8.8cm]{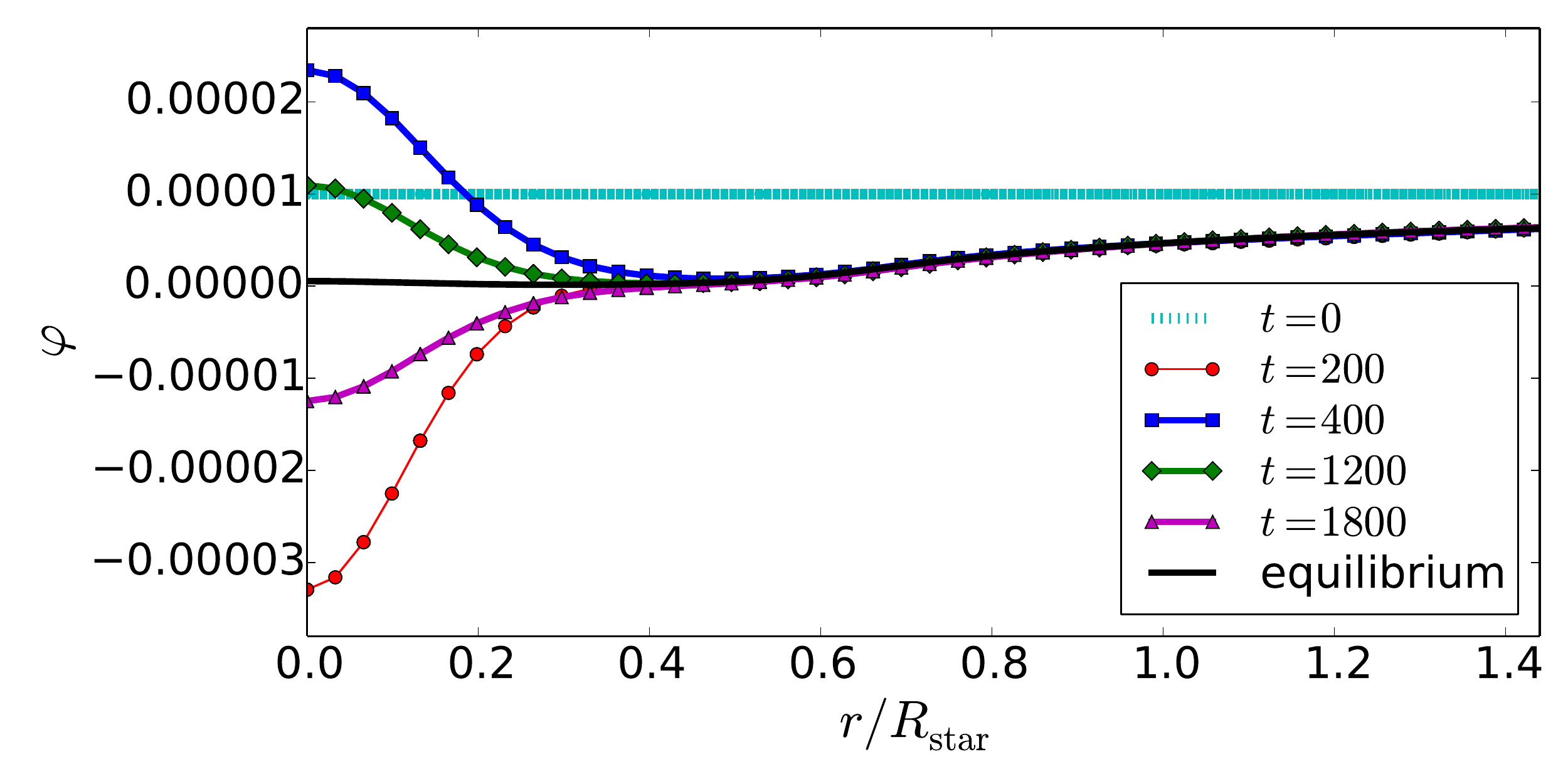}
\caption{{\it Evolution of the scalar field for a highly compact ($C=0.29$), non-rotating star
with $\beta=80$}.
The scalar field as a function of radius at different times during the evolution is shown.
Also shown is the equilibrium configuration (solid black) within ST with the same central density
as the initial data.
At late times the solution relaxes to the equilibrium configuration with a slowly damped, oscillating
central region.
\label{fig:C29_radial_profile}}
\end{figure}

These evolutions suggest that for small $\beta$, the configurations evolve to a stable, ST
equilibrium solution. Because the scalar field is the most dynamic aspect of the solution
in this regime, we plot it for a few times as it settles to its final state in 
Fig.~\ref{fig:C29_radial_profile}. Included in the figure is the corresponding static
star found by integrating the analogous TOV-like system of equations. The evolution
approaches the static star oscillating with a damped amplitude within the central region.

\begin{figure}
\centering
\includegraphics[width=8.8cm]{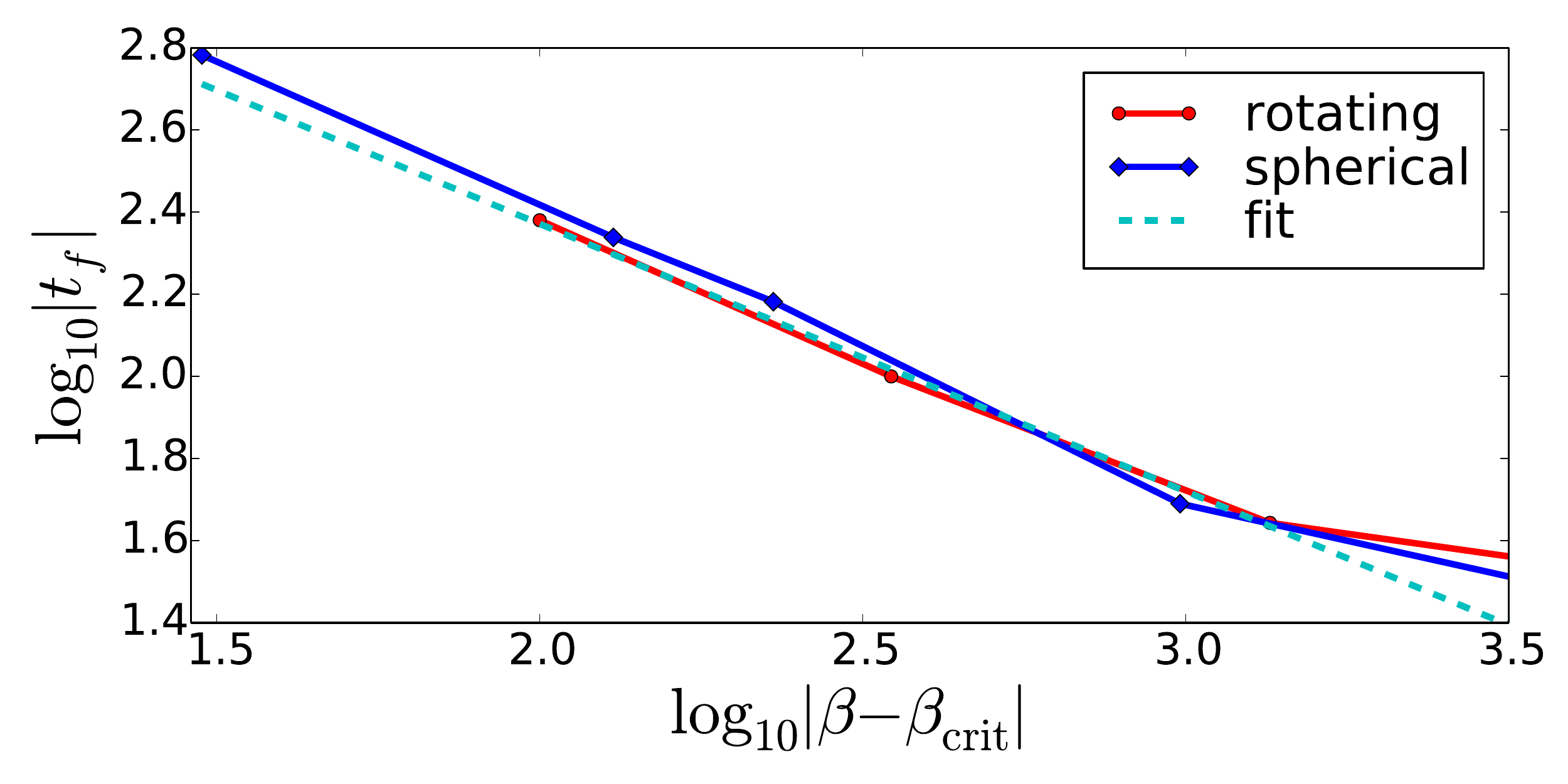}
\caption{{\it Growth rate for highly compact ($C=0.29$) stars}.
To determine the growth rate of the instability as a function of
$\beta$, the time, $t_f$, is found at which the scalar field
at the center has increased by a factor of $600$ (an arbitrary, fiducial value).
This fiducial time appears to scale as $t_f \propto |\beta-\beta_{\rm crit}|^\gamma$ with $\gamma \approx -0.65$ (cyan, dashed) for both non-rotating, spherical
stars (blue, solid) and rotating stars (red, solid).
%
\label{fig:growthrate}}
\end{figure}

For larger $\beta$, the instability drives
the evolution to collapse. As one decreases $\beta$ and approaches $\beta_{\rm crit}$ from
above, the growth of the central density is delayed further in time. One can estimate
the survival time of the configuration by choosing a large, fiducial value of the scalar field
and measuring the time it takes the central scalar field to reach this value.  This fiducial
time $t_f$ is plotted as a function of $\beta$ in Fig.~\ref{fig:growthrate}. It is notable
that both spherical and rotating solutions (with frequency $400$Hz) appear to demonstrate
the same scaling, namely that $t_f \propto |\beta-\beta_{\rm crit}|^{-0.65}$,
and that the scaling exponent, $-0.65$, is roughly comparable to the 
approximate value $-0.5$ presented in Ref.~\cite{1997PThPh..98..359H} [see Eq.~(5.3)].

\begin{figure}
\centering
\includegraphics[width=8.8cm]{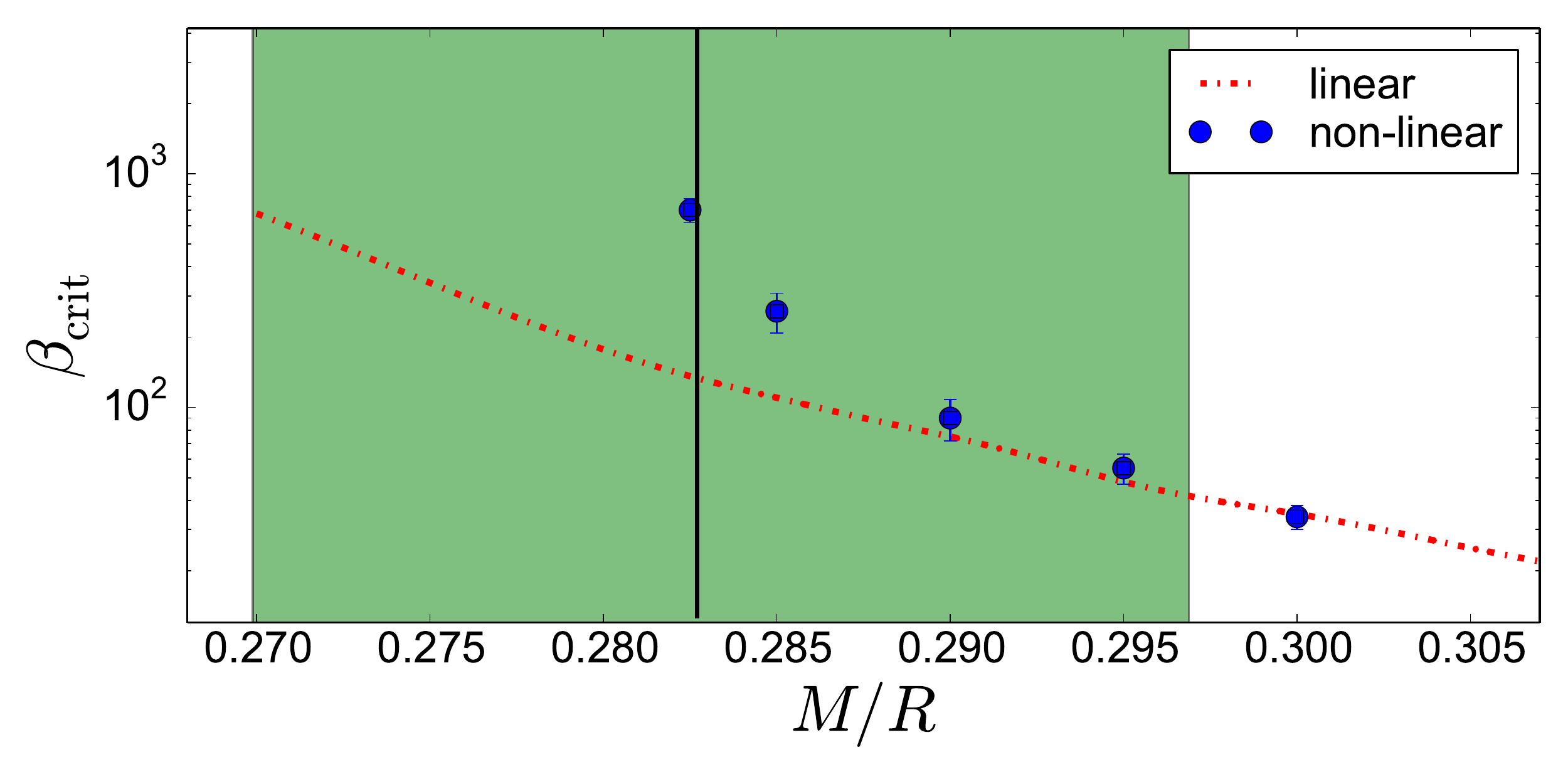}
\caption{{\it Stability region as a function of compactness for nonrotating solutions}.
The value of $\beta$ at which our numerical evolutions collapse to black hole are shown (blue dots)
along with a rough estimate of its uncertainty.
The results from the {\tt ENG} case of Ref.~\cite{2015PhRvD..91f4024M} (because of the similarity of
that case with the $\Gamma=3$ case) are also shown (red dotted curve) for comparison.
A vertical line (black solid line) indicates the compactness of the most massive
star observed assuming a radius of $10.5$km~\cite{2015arXiv150505155O}. The shaded region (green)
around that vertical line indicates the uncertainty in radius
(i.e., assuming half of the maximum error) leading to a range
of $10-11$km.
\label{fig:betac}}
\end{figure}

For $C=0.28$ we find an intermediate regime in $\beta$. For such cases, we see the
growth of the instability in the center of the star. However, this growth ceases after which the
central region oscillates with an exponentially
decreasing
amplitude. The solution appears to settle to its corresponding ST equilibrium solution. The difference
between this case and the small $\beta$ cases would appear to be the initial instability in the central
region. The oscillations in this region propagate via the scalar degree of freedom to large radius.
The nature of the transition from this intermediate regime to the collapsing regime and the question
of whether
this intermediate regime occurs in the rotating case require further investigation.

~\\
\noindent\textbf{\textit{Conclusions}\label{conclusions}}
%
Our simulations of the fully nonlinear system find that very compact stars containing
a central region with $T>0$ become unstable for sufficiently large $\beta$, consistent with
the linear perturbation results.
This instability, studied in detail in~\cite{2015PhRvD..91f4024M} in spherical symmetry
for different
microphysical EoS, appears for values of
$\beta_{\rm crit} \approx 140$ for a compactness of $C=0.283$. 
However, our evolutions show
that prompt collapse occurs for $\beta >700$, and so
establishing constraints from observational data may require
fully nonlinear solutions. 
Fig.~\ref{fig:betac} displays this difference 
by showing the values of $\beta$ for which the neutron star undergoes
prompt collapse together with the $\beta_{\rm crit}$ obtained from the linear analysis.
In addition, our evolutions extend to rotating solutions which break spherical symmetry and which
also demonstrate this same instability.
 
\begin{figure}
\centering
\includegraphics[width=8.8cm]{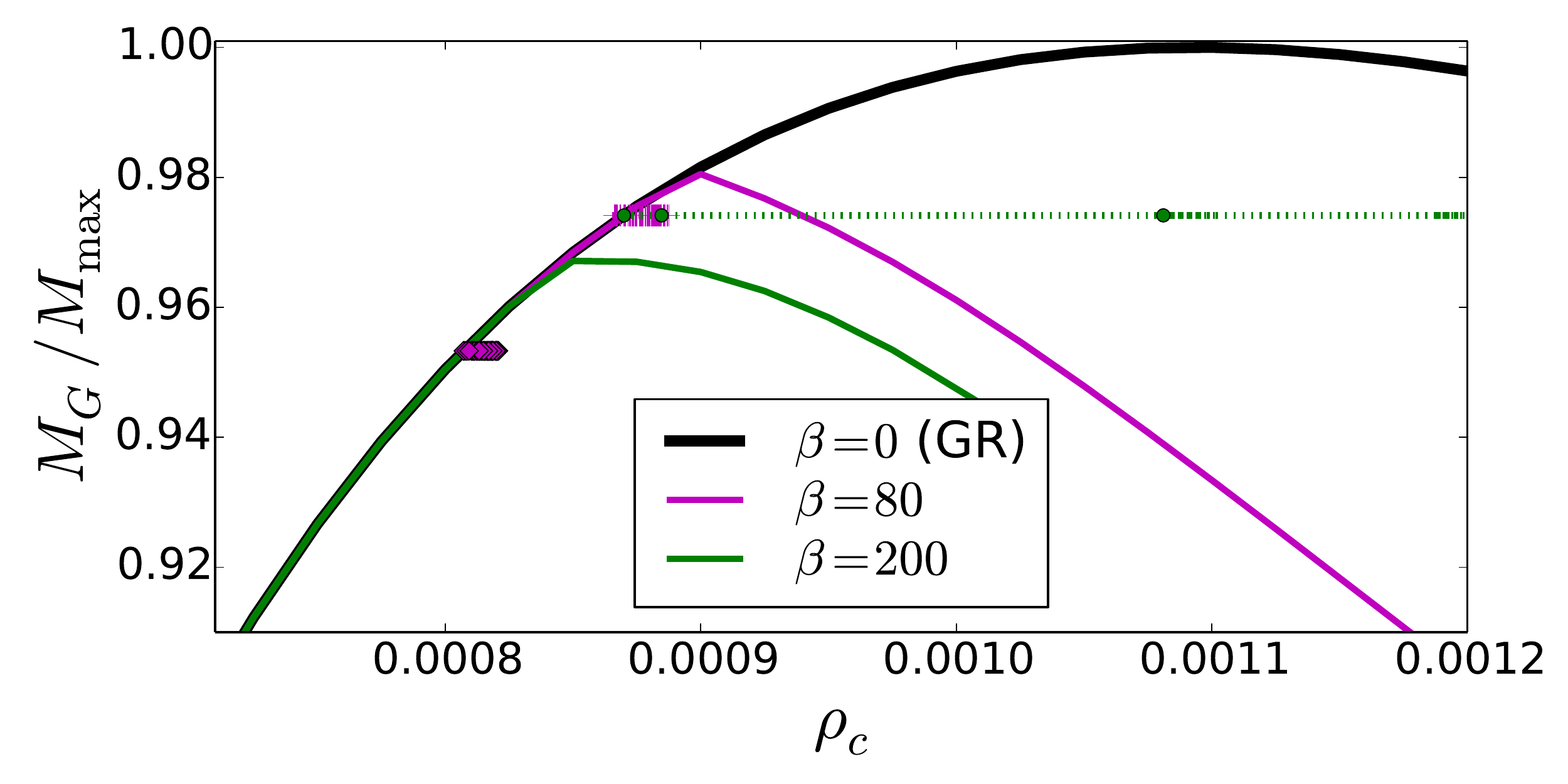}
\caption{{\it Sketch of the configuration space}.
Three curves show families of  equilibrium solutions for $\beta=0$ (GR), $\beta=80$,
and $\beta=200$. Two GR solutions are shown representing the
initial data used (along with an initially constant scalar field).
The evolution of a low compactness star (the lower solution) shows that the solution
oscillates about the nearly identical ST solution, regardless of $\beta$.
In contrast, the very compact
star (with larger central density) has two very different outcomes.
For $\beta=80$, the evolution oscillates about the ST solution (magenta).
For $\beta=200$, however,
the evolution (green dashed) proceeds to the right and the solution collapses without approaching 
any equilibrium solutions.
\label{fig:phase}}
\end{figure}

An important question left unanswered by the perturbation analysis concerns the end-state
of the instability. Our evolutions find that these compact stars are driven either to apparently
stable ST equilibrium configurations with non-constant profiles for the scalar field
or to black hole formation (see Fig.~\ref{fig:phase}). In particular, we conjecture that when starting with a GR solution
of some particular mass that it will be driven to a ST solution if such a solution on the
stable branch exists with roughly the same mass (or  a bit less). However, for large $\beta$,
such an approach to stable solutions is more difficult
since the maximum mass gets smaller as $\beta$ increases (see, e.g., Fig.~4 of Ref.~\cite{2015PhRvD..91f4024M}).

There are at least a couple of different ways of considering the astrophysical implications
of this instability. Consider 
the most massive neutron star ever observed, which 
has mass $M_{\rm NS}= 2.01 \pm 0.04 M_{\odot}$~\cite{2013Sci...340..448A}. To compute its
compactness, consider different estimates which
favor a radius for  neutron stars of roughly
$R_{\rm NS} \approx 10.5^{+ 1.2}_{- 1.0}$km~\cite{2015arXiv150505155O} (see also~\cite{2015arXiv150507436C})
for this mass,  which would
indicate a compactness of $0.283 \pm 0.030$. Such a value places it within the unstable regime, but
we acknowledge a number of uncertainties.  In particular, the uncertainty in the compactness is
quite large. Significant uncertainty arises from our ignorance about
 the appropriate stellar EoS. Additionally, very long evolutions of some cases
eventually collapse, and ensuring that the finite boundary is not affecting the solution
is difficult. Such a computational difficulty leads to uncertainty in the precise
value of $\beta_{\rm crit}$.

Perhaps, if gravity is described by DEF ST gravity, the existence of this extreme
star argues for an upper limit on $\beta$. If $\beta$
were higher, such a star might become unstable to collapse. Of course, a precise value
of our $\beta$ awaits more detailed knowledge of the star, but 
very likely $\beta$ would be bounded from above by $\beta_{\rm crit} \approx \mathcal{O}(1000)$.

However, another interpretation, equivalent to that above, is that the maximum observed
NS mass is not set by the maximum supported by the particular EoS describing
neutron star matter. Instead, the maximum mass may be set by the onset of this
instability in ST gravity (or in related, alternative theories such as $f(R)$ gravity~\cite{Capozziello:2015yza}). Future observations of massive neutron stars may elucidate this ambiguity and allow for more stringent constraints. 

\textbf{\textit{Acknowledgments:}}
We thank Enrico Barausse, Raissa Mendes, and Paolo Pani for helpful
discussions. 
We also acknowledge the support and hospitality of the
Fields Institute during their focus program \textit{100 Years of General
Relativity} where this work began.
This work was supported by the NSF under grant PHY-1308621~(LIU)
and by NASA's ATP program through grant NNX13AH01G.
CP acknowledges support from the Spanish Ministry of Education and Science through a Ramon y Cajal grant and from the Spanish Ministry of Economy and Competitiveness grant FPA2013-41042-P. Computations were performed at XSEDE, Scinet,
and MareNostrum.

\bibliographystyle{utphys}
\bibliography{paper}

\end{document}